\documentclass[final,square,comma,sort&compress,3p,time,twocolumn,groupedaddress,showpacs,floatfix]{elsarticle}
\bibpunct{[}{]}{,}{n}{,}{,}

\usepackage{graphics}

\journal{Physics Letter B}

\begin{document}

\begin{frontmatter}


\title{The mutable nature of particle-core excitations with spin\\  
in the one-valence-proton nucleus $^{133}$Sb}



\author{G. Bocchi$^{a,b}$}
\author{S. Leoni$^{a,b}$\corref{cor1}}
\author{B. Fornal$^{c}$}
\author{G. Col\`o$^{a,b}$}
\author{P.F. Bortignon$^{a,b}$}
\author{S. Bottoni$^{a,b}$}
\author{A. Bracco$^{a,b}$\\}
\author{C. Michelagnoli$^{d}$}
\author{D. Bazzacco$^{e}$}
\author{A. Blanc$^{f}$}
\author{G. De France$^{d}$}
\author{M. Jentschel$^{f}$}
\author{U. K\"{o}ster$^{f}$, \\P. Mutti$^{f}$}
\author{J.-M. R\'egis$^{g}$}
\author{G. Simpson$^{h}$}
\author{T. Soldner$^{f}$}
\author{C.A. Ur$^{e,i}$}
\author{W. Urban$^{j}$}
\author{L.M. Fraile$^{k}$, \\R. Lozeva$^{l}$}
\author{B. Belvito$^{a,b}$}
\author{G. Benzoni$^{b}$}
\author{A. Bruce$^{m}$}
\author{R. Carroll$^{n}$}
\author{N. Cieplicka-Ory\`nczak$^{b,c}$}
\author{F.C.L. Crespi$^{a,b}$, \\F. Didierjean$^{l}$}
\author{J. Jolie$^{g}$}
\author{W. Korten$^{o}$}
\author{T. Kr\"{o}ll$^{p}$}
\author{S. Lalkovski$^{n,q}$}
\author{H. Mach$^{k}$}
\author{N. M\u{a}rginean$^{r}$, \\B. Melon$^{s}$}
\author{D. Mengoni$^{e,t}$}
\author{B. Million$^{b}$}
\author{A. Nannini$^{s}$}
\author{D. Napoli$^{u}$}
\author{B. Olaizola$^{k}$}
\author{V. Paziy$^{k}$, \\Zs. Podoly\'ak$^{n}$}
\author{P.H. Regan$^{n,v}$}
\author{N. Saed-Samii$^{g}$}
\author{B. Szpak$^{c}$}
\author{V. Vedia$^{k}$}


\address{
$^a$ Dipartimento di Fisica, Universit$\grave{a}$ degli Studi di Milano, I-20133 Milano, Italy\\
$^b$ INFN sezione di Milano via Celoria 16, 20133, Milano, Italy \\
$^c$ Institute of Nuclear Physics, PAN, 31-342 Krak\'ow, Poland \\
$^d$ GANIL, BP 55027, 14076 Caen CEDEX 5, France\\
$^e$ INFN Sezione di Padova, I-35131 Padova, Italy\\
$^f$ ILL, 71 Avenue des Martyrs, 38042 Grenoble CEDEX 9, France\\
$^g$ IKP, University of Cologne, Z\"{u}lpicher Str. 77, D-50937 K\"{o}ln, Germany\\
$^h$ LPSC, 53 Avenue des Martyrs, F-38026 Grenoble, France\\
$^i$ ELI-NP Magurele-Bucharest, Romania\\
$^j$ Faculty of Physics, Warsaw Univ., ul. Hoza 69, PL-00-681 Warsaw, Poland\\
$^k$ Grupo de F\'isica Nuclear, Universidad Complutense, CEI Moncloa, 28040
Madrid, Spain\\
$^l$ IPHC, CNRS/IN2P3 and University of Strasbourg, F-67037 Strasbourg, France\\
$^m$ SCEM, University of Brighton, Lewes Road, Brighton BN2 4GJ, UK\\
$^n$Department of Physics, University of Surrey, Guildford, GU2 7XH, UK\\
$^o$ CEA, Centre de Saclay, IRFU, F-91191 Gif-sur-Yvette, France\\
$^p$ Institut f\"{u}r Kernphysik, TU Darmstadt, Schlossgartenstrasse 9 64289 Darmstadt, Germany\\
$^q$ Faculty of Physics, University of Sofia, 5 James Bourchier Blvd, 1164 Sofia,
Bulgaria\\
$^r$ Horia Hulubei National Institute of Physics and Nuclear Engineering - IFIN HH, Bucharest, 077125, Romania\\
$^s$ INFN Sezione di Firenze, Firenze, Italy \\
$^t$ Dipartimento di Fisica, Universit$\grave{a}$ degli Studi di Padova, I-35131 Padova, Italy \\
$^u$ INFN, Laboratori Nazionali di Legnaro, I-35020 Padova, Italy \\
$^v$ Acoustics and Ionizing Radiation Division, National Physical Laboratory,
Teddington, Middlesex, TW11 0LW, UK }


\date{\today}

\begin{abstract}
The $\gamma$-ray decay of excited states of the one-valence-proton nucleus $^{133}$Sb has been studied using cold-neutron induced fission of $^{235}$U and $^{241}$Pu targets, during the EXILL campaign at the ILL reactor in Grenoble. By using a highly efficient HPGe array, coincidences between $\gamma$-rays prompt with the fission event and those delayed up to several tens of microseconds were investigated, allowing to observe, for the first time, high-spin excited states above the 16.6 $\mu$s isomer. Lifetimes analysis, performed by fast-timing techniques with LaBr$_3$(Ce) scintillators, reveals a difference of almost two orders of magnitude in B(M1) strength for transitions between positive-parity medium-spin yrast states. The data are interpreted by a newly developed microscopic model which takes into account couplings between core excitations (both collective and non-collective) of the doubly magic nucleus $^{132}$Sn and the valence proton, using the Skyrme effective interaction in a consistent way. The results point to a 
fast change in the nature of particle-core excitations with increasing spin.
\end{abstract}


\cortext[cor1]{Corresponding author: silvia.leoni@mi.infn.it}

\end{frontmatter}

The structure of atomic nuclei can be viewed from two general and complementary perspectives: a microscopic one, focusing on the motion of individual nucleons in a mean field potential created by all constituents, giving rise to the quantum shell structure, and a mesoscopic perspective that focuses on a highly organized complex system, exhibiting collective behavior. Ideal systems to investigate this duality should be nuclei composed of one valence particle and a doubly magic core in which the coupling between collective core excitations (phonons) and the valence nucleon strongly influences the structure of the system \cite{BM}. The understanding of this particle-phonon coupling is of primary importance, being responsible for the anharmonicities of vibrational spectra \cite{BM,Sol95}, the quenching of spectroscopic factors \cite{Pan97,Gad08,Bar09,Tsa09,All14} and the reduction of $\beta$-decay half-lives in magic nuclei \cite{Niu15}; it is also the key process at the origin of the damping of giant resonances \cite{Bor98}. In general, the coupling between phonons and particles is at the basis of fermionic many-body interacting systems, both in nuclear physics and in condensed matter physics \cite{Colo04}.

In reality, in nuclear physics a more complex scenario is realized: collective phonons are not the only excitations at low energy in doubly magic systems -  usually states having real phonon character co-exist here with excitations that are less collective or have no collective properties. In this respect, a benchmark region is around $^{132}$Sn, which is one of the best doubly magic cores and exhibits low-lying both collective and non-collective excitations. In the present work, we had the goal of investigating the nature of particle-core excitations in the corresponding one-valence-proton nucleus $^{133}$Sb, populated in cold-neutron induced fission. First, we aimed at identifying, experimentally, new high spin yrast states above the long-lived 16.6 $\mu$s isomer. This required a demanding technique which relies on measuring coincidences between $\gamma$ rays prompt with the fission event and those delayed up to several tens of microseconds. Then, we studied transition probabilities through lifetime measurements of selected states. To interpret the data, a new microscopic and self consistent model has been developed, containing particle couplings to core excitations of various nature: in such a heavy mass region this cannot be treated with
a shell model (SM) approach as it would require full SM calculations in
the configuration space that encompasses proton and neutron orbitals below and above $^{132}$Sn
\cite{Lac09,SciDAC07}. We anticipate that experimental data on $^{133}$Sb, in the light of model predictions, provide evidence for a fast change in the nature of particle-core excitations, from a collective character to a non-collective one with increasing spin. 

So far, studies of particle-core excitations considered, almost exclusively, couplings with collective phonons of the core. The most known case is the multiplet in $^{209}$Bi (one-proton nucleus with respect to the $^{208}$Pb core) arising from the coupling of a h$_{9/2}$ proton with the 3$^-$ phonon of $^{208}$Pb (at 2615 keV), exhibiting one of the largest vibrational collectivity across the nuclear chart (34 W.u.) \cite{BM}. In other one-particle (1p) or one-hole (1h) nuclei around $^{208}$Pb \cite{Pie09, Rej00, Wil15} and other magic nuclei \cite{Kle82, Lun84, Gal88, Lis80, Mon11, Mon12, Boc14, Nit14}, states originating from couplings of the 3$^-$ phonon with single particle/hole have been located as well. In the past, the theoretical description of particle-phonon couplings relied on phenomenological models \cite{BM,Sol95}. Now, microscopic approaches based on either Skyrme forces or Relativistic Mean Field (RMF) Lagrangians are feasible, but with applications limited to the description of single particle states \cite{Col10,Cao14,Lit06,Lit11} and giant resonances \cite{Niu14,Lit14}.

In the case of the $^{132}$Sn core, the first three excitations, 2$^+$ at 4041 keV, 3$^-$ at 4352 keV and 4$^+$ at 4416 keV show a less pronounced collectivity (of the order of 7 W.u.) with respect to the 3$^-$ of $^{208}$Pb, and the other states have 1p-1h character \cite{Nud2,Bat01}. In consequence, the one-valence-proton nucleus $^{133}$Sb, being bound up to 7.4 MeV (unlike $^{133}$Sn with a neutron binding energy of only 2.4 MeV), is a perfect case to test, simultaneously, the coupling of a particle with core excitations of various nature.

An extension of $\gamma$-spectroscopy of $^{133}$Sb is challenging: so far, all states having a single proton on one of the d$_{5/2}$, h$_{11/2}$, d$_{3/2}$ and s$_{1/2}$ orbitals, as dominant configuration, have been located. Also, a series of excitations with J$^{\pi}$ = 7/2$^+$,  9/2$^+$,  11/2$^+$,  13/2$^+$,  15/2$^+$,  17/2$^+$ and 21/2$^+$, of which the 21/2$^+$ state is isomeric with 16.6 $\mu$s half-life, are known from isomer and $\beta$-decay studies \cite{Urb00,Urb09,Gen00,San99}. On the contrary, no information exists on positive-parity levels above the long-lived 21/2$^+$ isomer, which is known with  x $<$ 30 keV uncertainty in energy \cite{Urb09,Sun10}. 


\begin{figure}[htc]
\resizebox{0.5\textwidth}{!}{\includegraphics{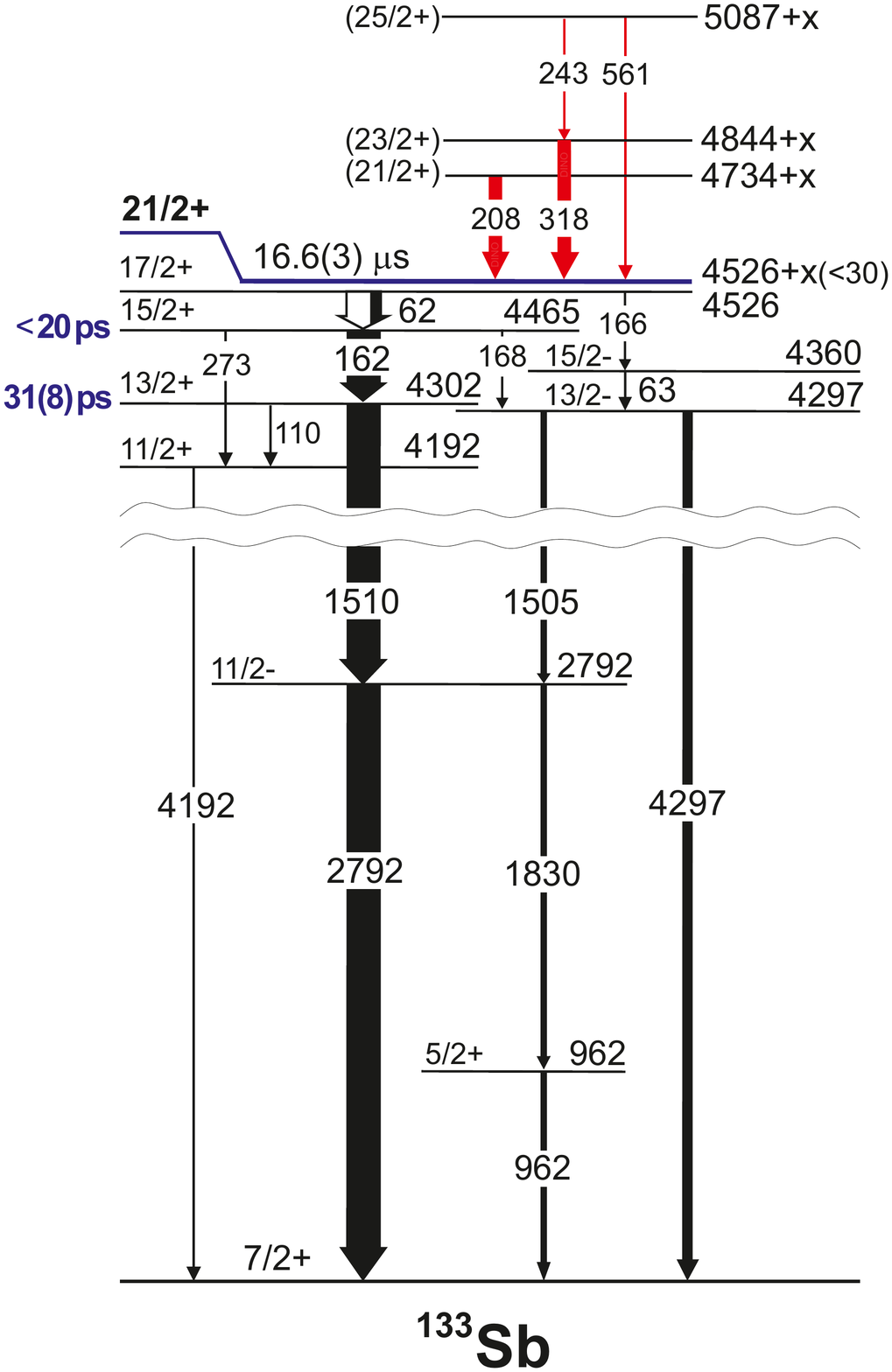}}
\caption{(Color online) Experimental level scheme of $^{133}$Sb: in black, the decay below the long lived 21/2$^+$ isomer, known prior to this work (being x $<$ 30 keV the isomer energy uncertainty)\cite{Urb09}; in red, newly identified transitions, above the isomer. The half lives of the 13/$2^+$ and 15/$2^+$ states - deduced from this work - are also given.} 
\label{level-scheme}
\end{figure}

The $\gamma$-ray coincidence data on $^{133}$Sb were obtained with a highly efficient HPGe array, installed at the PF1B \cite{Abe06} cold-neutron facility at Istitut Laue Langevin (Grenoble, France). The ILL reactor is a continuous neutron source with an in-pile flux up to 1.5$\times$10$^{15}$ neutrons cm$^{-2}$ s$^{-1}$. After collimation to a halo-free pencil beam,
the capture flux on target was 10$^{8}$ neutrons cm$^{-2}$ s$^{-1}$. Two detector setups were used, the first consisting of 8 EXOGAM clovers \cite{exogam}, 6 large coaxial detectors from GASP \cite{gasp} and 2 ILL-Clover detectors, with a total photopeak efficiency of about 6$\%$ at 1.3 MeV. In the second setup, the GASP and ILL detectors were replaced by 16 LaBr$_3$(Ce) detectors, named FATIMA array \cite{FATIMA}, for lifetime measurements by fast-timing techniques. This is the first time a large HPGe array has been installed around such a high intensity, highly collimated cold-neutron beam \cite{Mut13,DeF14,Jen15}.

The campaign, named EXILL, lasted two reactor cycles (each $\approx$50 days long) and its main part consisted of two long runs of neutron induced fission on $^{235}$U and $^{241}$Pu targets. The use of a fully digital, triggerless acquisition system (with time stamp intervals of 10 ns) allowed event rates up to 0.84 MHz to be handled and to study coincidences among $\gamma$ transitions separated in time by several tens of microseconds \cite{Leo15} -- with analogue electronics, coincidences only across a few $\mu$s isomers could be studied with large Ge arrays.

\begin{figure}[htc]
\resizebox{0.48\textwidth}{!}{\includegraphics{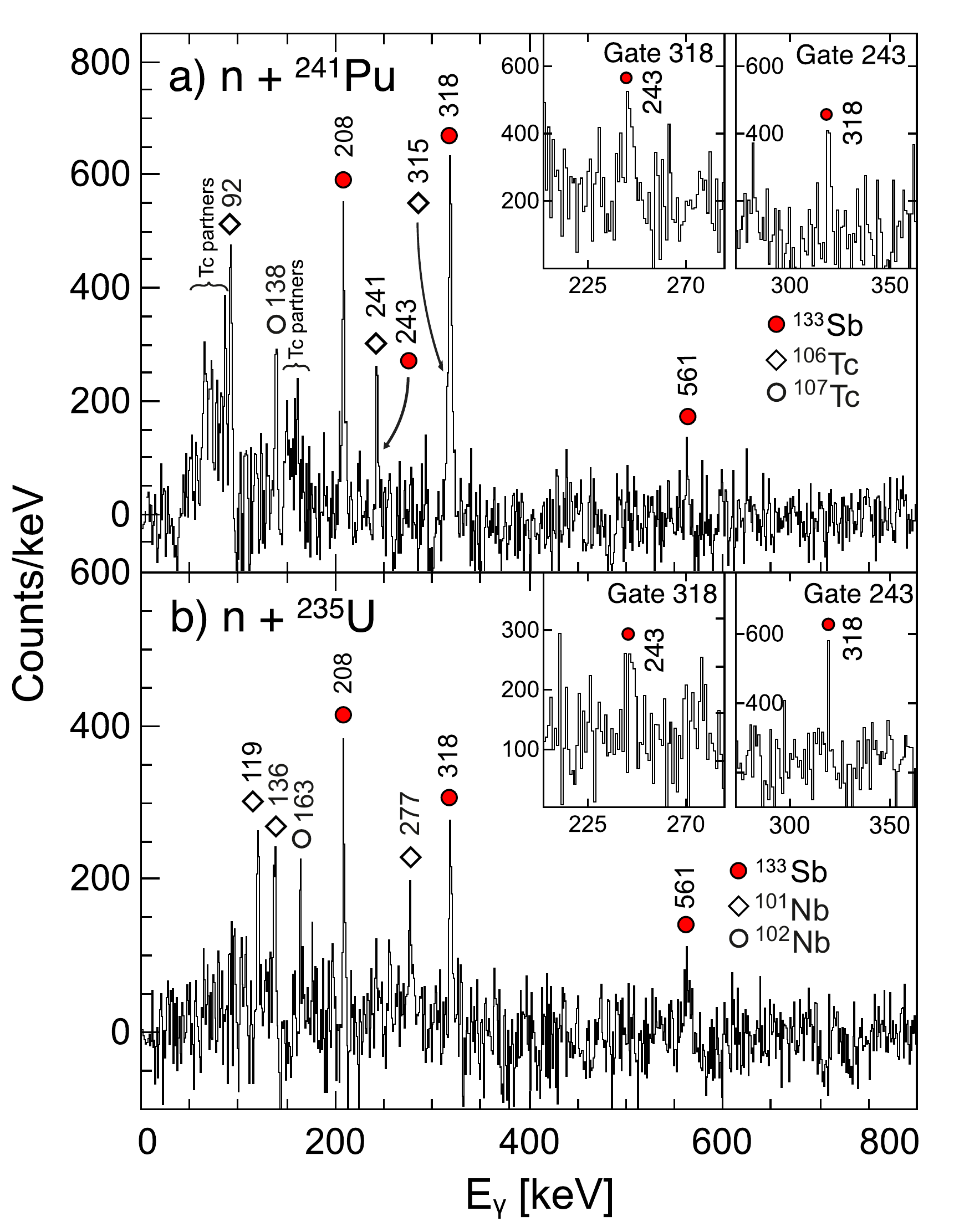}}
\caption{(Color online) Spectra of $^{133}$Sb $\gamma$-rays preceding the 16.6 $\mu$s isomer, obtained in cold-neutron induced fission of $^{241}$Pu (a) and $^{235}$U (b) targets, gated on pairs of delayed transitions depopulating the isomer. The insets show portions of spectra gated on 318 or 243 keV, in prompt coincidence with fission events and $\gamma$ rays deexciting the isomer (see Fig. \ref{level-scheme} and text for details). } 
\label{gamma-spectra}
\end{figure}

In $^{133}$Sb, the 21/2$^+$ isomeric state decays via a cascade of five transitions: an unknown isomeric transition with E$_{\gamma} < 30$ keV followed by 62, 162, 1510 and 2792 keV $\gamma$ rays that feed the 7/2$^+$ ground state (see Fig. \ref{level-scheme}). Therefore, a search for high spin structures of $^{133}$Sb was undertaken by considering coincidences between two classes of $\gamma$-rays: i) \textit{prompt} $\gamma$-rays - coincident (within 200 ns) with a fission event (defined by $\gamma$-ray multiplicity equal or larger than 4, within 200 ns) and ii) \textit{delayed} $\gamma$ rays - emitted  within 20 $\mu$s after the fission event and coincident (within 200 ns) with at least one of the four known transitions deexciting the 21/2$^+$ isomer. First, we investigated a \textit{prompt}-\textit{delayed} matrix. Fig. \ref{gamma-spectra} (a) and (b) show spectra of $\gamma$ rays preceding the 16.6 $\mu$s isomer, obtained from the $^{241}$Pu and $^{235}$U targets, respectively. The $\gamma$ rays observed in both data sets at 207.9(4), 318.0(4), and 561(1) keV are candidates for transitions occurring higher in the level scheme of $^{133}$Sb. In addition, by exploiting the \textit{prompt}-\textit{prompt} coincidence histogram, constructed in coincidence with a \textit{delayed} $\gamma$ ray deexciting the isomer, a new weak 243-keV line was identified in coincidence with the 318-keV transition (see insets of Fig. \ref{gamma-spectra}). This line was then placed in cascade with the 318-keV $\gamma$ ray, depopulating a level located at 561 keV above the isomer. This placement is strongly supported by the existence of a 561-keV transition that was found above the isomer. As the 318-keV transition has much higher intensity then the newly found 243-keV line, it can be placed as feeding the 21/2$^+$ isomer. In this way, we have located three new levels, with energies 4734+x, 4844+x and 5087+x keV, as shown in Fig. \ref{level-scheme}.

\begin{figure}[htc]
\resizebox{0.48\textwidth}{!}{\includegraphics{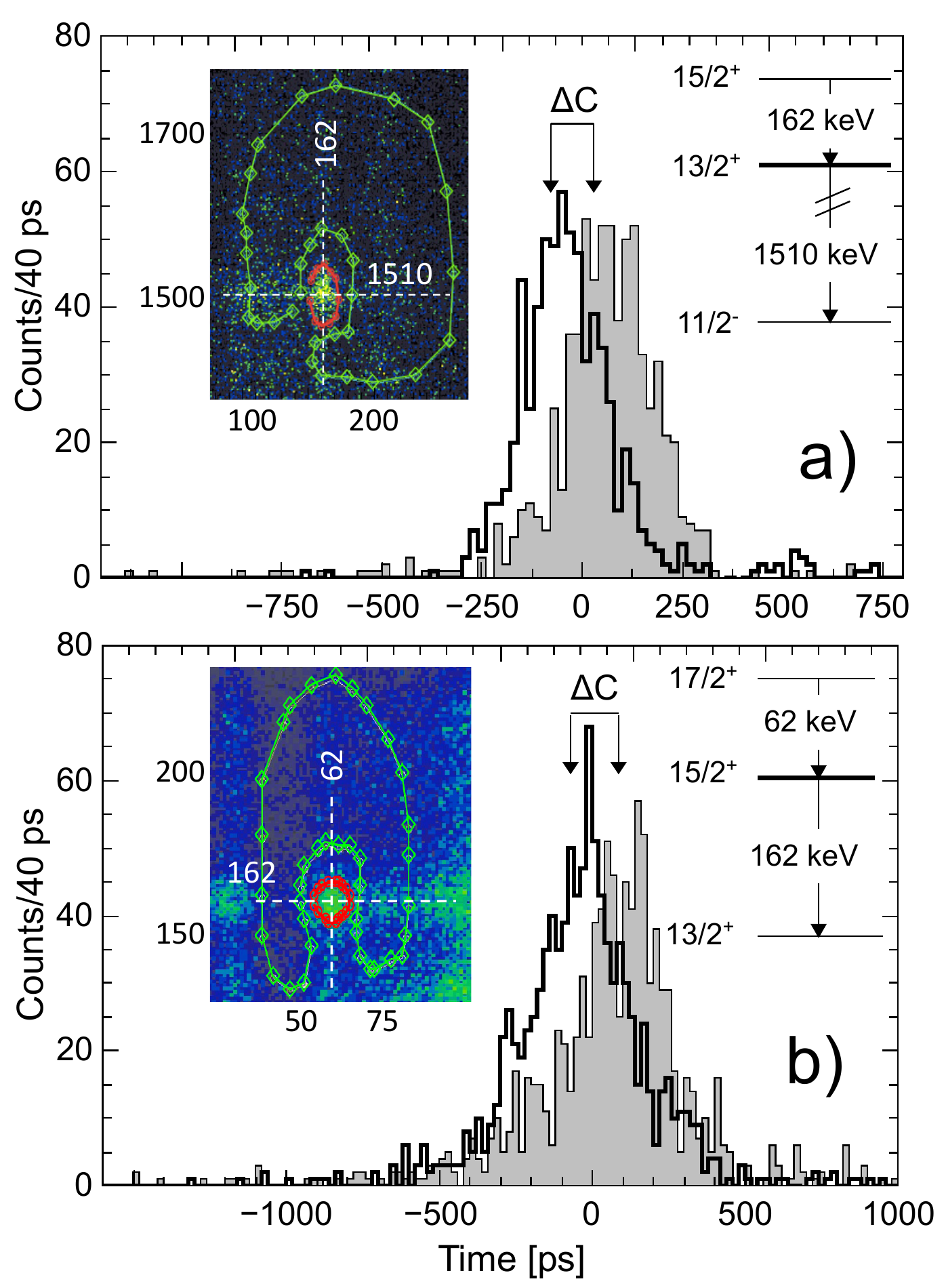}}
\caption{(Color online) Time spectra used for the lifetime analysis of the 13/2$^+$ (a) and 15/2$^+$ states (b) of $^{133}$Sb ($^{235}$U fission data). Dark (light) histograms are the distributions of the time difference between the detection of the feeding and decaying transitions (and vice versa), for the cascades shown on the right. $\Delta$C is the difference between the centroids. The insets show examples of background selections around the coincidence peaks in the (E$_{\gamma_1}$, E$_{\gamma_2}$) LaBr$_3$(Ce) histograms.} 
\label{timespectra}
\end{figure}

In the second part of this work, fission data from $^{235}$U and $^{241}$Pu targets, taken with the setup including the LaBr$_3$(Ce) scintillators, were used to extract the lifetimes of the 13/2$^+$ and 15/2$^+$ states of $^{133}$Sb, by fast-timing techniques \cite{Reg13,Reg14,Reg14b}. The analysis was based on triple coincidence events, within a time window of 200 ns, in which two $\gamma$-rays are detected in the LaBr$_3$(Ce) scintillators and the third one in the Ge array. By setting the very selective gate on the 2792-keV line of $^{133}$Sb recorded in the Ge, a (E$_{\gamma_1}$,E$_{\gamma_2}$,$\Delta$t) histogram was constructed, $\Delta$t being the time difference between $\gamma$ rays with E$_{\gamma_1}$, E$_{\gamma_2}$ energies, measured by the LaBr$_3$(Ce) detectors. Figure \ref{timespectra} shows the time distributions used in the analysis of the 13/2$^+$ (a) and 15/2$^+$ (b) states, respectively ($^{235}$U data set), and associated with the 
$\gamma$-ray cascades shown on the right \cite{Reg14}. All time spectra are background subtracted by considering a two-dimensional gate in the energy plane, around the corresponding coincidence peak (as shown in the insets). The time difference $\Delta$C between the centroids of the time distributions provides the lifetime $\tau$ of the state (2$\tau$ = $\Delta$C - PRD), after correction for the prompt response difference (PRD) \cite{Reg14}. 
From the $^{235}$U data, half-lives T$_{1/2}$ = 32(10) ps and $<$ 17 ps were deduced for the 13/2$^+$ and 15/2$^+$ states, respectively, in very good agreement with the 30(12) ps and $<$ 21 ps values obtained from the $^{241}$Pu data. This gives the average values T$_{1/2}$ = 31(8) ps and $<$ 20 ps. Taking into account the decay branchings from the two levels \cite{Urb09}, B(M1) values were extracted for the 15/2$^+ \rightarrow$13/2$^+$ and 13/2$^+ \rightarrow$11/2$^+$ transitions, yielding $>$ 0.24 W.u. and 0.0042(15) W.u., respectively. This large difference, of almost two orders of magnitude, is clearly intriguing and brings a signature of some non-trivial change of configuration mixing in the 11/2$^+$, 13/2$^+$ and 15/2$^+$ states of $^{133}$Sb, respectively.

In order to interpret the experimental findings, a new microscopic model has been developed with the aim of describing states with different degrees of collectivity. 
In our model we solve the Hamiltonian $H=H_0+V$ with
\begin{eqnarray}
H_0 & = & \sum_{j} \varepsilon_{j} a^\dagger_{j} a_{j} + \sum_{nJ} \hbar\omega_{nJ}
\Gamma^\dagger_{nJ} \Gamma_{nJ}, \nonumber \\
V & = & \sum_{jj'}\sum_{nJ} 
\frac{\langle j \vert\vert V \vert\vert j'nJ \rangle}{\sqrt{2j+1}}
a^\dagger_{j} \left[ a_{j'} \otimes \Gamma_{nJ} \right]_{j},
\end{eqnarray}
where we have written for simplicty $j$ instead of $nlj$, and the phonons 
with angular momentum $J$ are labelled by the index $n$. 
Our calculation has no free parameters and is self-consistent in the sense that both single-particle
states and phonons come out of Hartree-Fock (HF) and Random Phase Approximation (RPA) calculations performed with the Skyrme SkX interaction \cite{Bro98}.
This Hamiltonian can be diagonalized separately in different Hilbert subspaces with good angular momentum $j$ and parity $\pi$. 
In each of these subspaces we have both one-particle states and 
so-called one particle-one phonon states. However, some of the so-called
phonons turn out to be pure 1p-1h states, as it can be expected. 
We label all states as phonons just in keeping with the fact that
they come out of the RPA diagonalization, but the word ``excitation'' 
would indeed be more appropriate. 
Table \ref{table:phonons} shows the calculated excitations of the $^{132}$Sn core, in comparison with known experimental data; 
we presently include those up to 5.5 MeV, together with the proton states of the 50-82 shell. 
We call our model ``Hybrid Configuration Mixing'' (HCM) and 
its details, as well as a detailed account of the results, will
be published elsewhere. The key point is that the correction for the
non-orthonormality of the basis is taken into account \cite{Row69} by solving 
the generalized eigenvalue problem 
$\left( H - E_\lambda N \right) \vert \lambda \rangle = 0$, 
where $N$ is the overlap matrix between the basis states.

\begin{figure*}[htc]
\resizebox{1.0\textwidth}{!}{\includegraphics{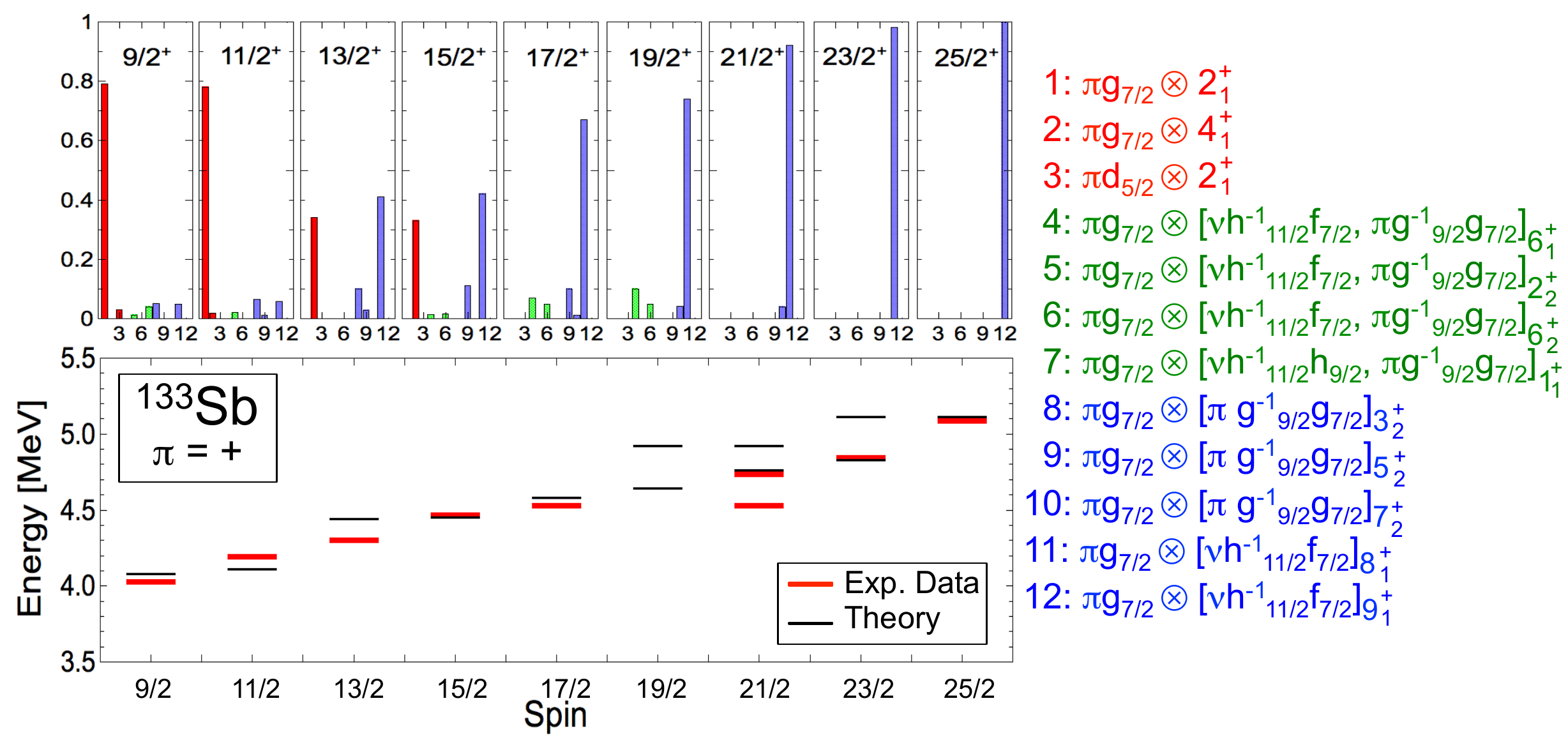}}
\caption{(Color online) Bottom panel: experimental and calculated energies (thick and thin lines, respectively) of the low-lying positive parity states of $^{133}$Sb, in the spin range 9/2$^+$ - 25/2$^+$. In the calculations, other states below 5.5 MeV are shown at 19/2$^+$, 21/2$^+$ and 23/2$^+$. Top panel: the components of each lowest state (with amplitude larger than 0.01), with configurations indicated in the legend. Lifetimes for the 13/2$^+$ and 15/2$^+$ states were measured in this work.} 
\label{fragmentation}
\end{figure*}

Figure \ref{fragmentation} shows, in the bottom panel, the calculated yrast and near yrast states of $^{133}$Sb, arising from the coupling between the valence proton and core-excitations. A number of features is evident. The model does reproduce well the energy sequence of the high-spin states 
observed experimentally. Also, a fast evolution of the wave function composition is seen, 
from complex to non-collective character, with increasing spin. As shown in the top panels, the low spin states are dominated by the g$_{7/2}$ proton coupled to the 2$^+$ phonon, while the highest spin excitations arise mostly from this valence proton coupled to the neutron h$^{-1}_{11/2}$f$_{7/2}$ non-collective core excitation. The states in between, at spin 13/2$^+$ and 15/2$^+$, show instead a fragmented wave function involving the coupling of the valence proton to both the 4$^+$ phonon and non-collective particle-hole excitations. 

\begin{table}[hbt]
\resizebox{\linewidth}{!}{
\begin{tabular}{l|ll|ll|l}
\hline
I$^{\pi}$& \multicolumn{2}{|c}{Energy [MeV]} & \multicolumn{2}{|c|}{B(E/M$\lambda$) [W.u.]}
& Main components \\
& Exp. & Theory & Exp. & Theory & Theory \\
\hline
2$^+_1$ & 4.041   & 3.87  & 7 & 4.75 &
$\nu {\rm h}_{11/2}^{-1}{\rm f}_{7/2}$ (0.56), 
$\pi {\rm g}_{9/2}^{-1}{\rm d}_{5/2}$ (0.19), \\
& & & & &
$\pi {\rm g}_{9/2}^{-1}{\rm g}_{7/2}$ (0.14) \\
3$^-_1$ & 4.352   & 5.02  & $>$ 7.1 & 9.91 & 
$\nu {\rm s}_{1/2}^{-1}{\rm f}_{7/2}$ (0.40),
$\nu {\rm d}_{3/2}^{-1}{\rm f}_{7/2}$ (0.12), \\
& & & & &
$\pi {\rm p}_{1/2}^{-1}{\rm g}_{7/2}$ (0.12)
\\
4$^+_1$ & 4.416   & 4.42 & 4.42 & 5.10 & 
$\nu {\rm h}_{11/2}^{-1}{\rm f}_{7/2}$ (0.63),          
$\pi {\rm g}_{9/2}^{-1}{\rm g}_{7/2}$ (0.21)
\\
6$^+_1$ & 4.716   & 4.73  && 1.65 & 
$\nu {\rm h}_{11/2}^{-1}{\rm f}_{7/2}$ (0.86),
$\pi {\rm g}_{9/2}^{-1}{\rm g}_{7/2}$ (0.11)
\\
8$^+_1$ & 4.848   & 4.80 && 0.28 &
$\nu {\rm h}_{11/2}^{-1}{\rm f}_{7/2}$ (0.98)
\\
5$^+_1$ & 4.885   & 4.77  && 0.20 &
$\nu {\rm h}_{11/2}^{-1}{\rm f}_{7/2}$ (0.99)
\\
7$^+_1$ & 4.942   & 4.80  && 0.30 & 
$\nu {\rm h}_{11/2}^{-1}{\rm f}_{7/2}$ (0.98)
\\
(9$^+_1$) & 5.280 & 4.99  && 0.04 &
$\nu {\rm h}_{11/2}^{-1}{\rm f}_{7/2}$ (0.99)
\\
1$^+_1$ &         & 4.97  && 7.95  &
$\pi {\rm g}_{9/2}^{-1}{\rm g}_{7/2}$ (0.76), 
$\nu {\rm h}_{11/2}^{-1}{\rm h}_{9/2}$ (0.24)
\\
2$^+_2$ &         & 5.37  && $<$ 10$^{-2}$  & 
$\pi {\rm g}_{9/2}^{-1}{\rm g}_{7/2}$ (0.72),
$\nu {\rm h}_{11/2}^{-1}{\rm f}_{7/2}$ (0.18) \\
2$^-_1$ &         & 5.44  && 0.47 & 
$\nu {\rm d}_{3/2}^{-1}{\rm f}_{7/2}$ (0.79)
\\
3$^+_1$ &         & 4.79  && 0.13  &
$\nu {\rm h}_{11/2}^{-1}{\rm f}_{7/2}$ (0.96) \\
3$^+_2$ &         & 5.40  && 1.99  &
$\pi {\rm g}_{9/2}^{-1}{\rm g}_{7/2}$ (0.96) \\
4$^+_2$ &         & 5.25  && 1.01  & 
$\pi {\rm g}_{9/2}^{-1}{\rm d}_{3/2}$ (0.56),
$\nu {\rm h}_{11/2}^{-1}{\rm f}_{7/2}$ (0.32) \\
5$^+_2$ &         & 5.45  &&  0.61 &
$\pi {\rm g}_{9/2}^{-1}{\rm g}_{7/2}$ (0.99) \\
6$^+_2$ &         & 5.32  &&  2.67 &             
$\pi {\rm g}_{9/2}^{-1}{\rm g}_{7/2}$ (0.74),
$\nu {\rm h}_{11/2}^{-1}{\rm f}_{7/2}$ (0.13) \\
7$^+_2$ &         & 5.42  &&  0.50  &
$\pi {\rm g}_{9/2}^{-1}{\rm g}_{7/2}$ (0.99) \\
\hline
\end{tabular}
}
\caption{Experimental and RPA multipole states of
$^{132}$Sn. The main components are those associated with RPA
amplitudes $X$ that are larger than 0.3 in absolute value, and
are listed together with the value of $X^2$ in parenthesis. 
\label{table:phonons}}
\end{table}

The states located in the present work above the 21/2$^+$ isomer, at 4.844+x and 5.087+x MeV (being x$<$30 keV), clearly correspond to the excitations calculated at 4.83 and 5.11 MeV. As seen in Fig. \ref{fragmentation}, they arise from almost pure ($>$95\%) configurations of $\pi$g$_{7/2}$ $\nu$f$_{7/2}$h$_{11/2}^{-1}$ character. One has to note that the existence of two yrast states above the 21/2$^+$ isomer with spin-parity assignments of 23/2$^+$ and 25/2$^+$ were suggested on the basis of the shell-model calculations with adjusted empirical interactions by W. Urban et al. \cite{Urb00,Urb09}. Similar sequences of states involving neutron particle-hole $\nu$f$_{7/2}$h$_{11/2}^{-1}$ excitations of the $^{132}$Sn core have been identified above 4 MeV in the neighboring nuclei $^{134}$Sb, $^{134}$Te and $^{135}$Te \cite{Zha96,For01}. A characteristic feature of these multiplets is that their members are often connected by M1 and E2 competing transitions. The decay of a level located at 5.087 MeV, and assigned as the highest spin member 25/2$^+$ of the $\pi$g$_{7/2}$ $\nu$f$_{7/2}$h$_{11/2}^{-1}$ configuration, has M1 and E2 branches which supports the suggested spin-parity assignments.

Concerning the two M1 transition probabilities that have been measured in the present work, theory
provides the values of 0.021 W.u. and 0.001 W.u. in the case of 15/2$^+$ $\rightarrow$ 13/2$^+$ and 13/2$^+$ $\rightarrow$ 11/2$^+$, 
respectively. The large ratio of $\approx$ 20 between the two values is in qualitative agreement with the experimental value 
of $\approx$ 60, and can be well understood from the point of view of our model, while it would not come out within the simple picture of Refs. \cite{Urb00,Urb09}. The 15/2$^+$ and 13/2$^+$ states have very similar composition of the wave function, and the largest component is the 2p-1h 
configuration ($\pi$g$_{7/2}$ $\nu$h$_{11/2}^{-1}$f$_{7/2}$) which has an amplitude of the order of 0.4. The transition matrix element 
$\langle 13/2^+ \vert\vert O(M1) \vert\vert 15/2^+ \rangle$, that is 0.78 $e\mu_N$, would become 4.86 $e\mu_N$ if a pure ($\pi$g$_{7/2}$ $\nu$h$_{11/2}^{-1}$f$_{7/2}$) component were assumed for both states, as in the 
simplified shell model of Refs. \cite{Urb00,Urb09}, leading to a B(M1) of 0.78 W.u. If we further assumed the same purity for the 11/2$^+$ state, the transition 
matrix element would remain approximately the same with an associated B(M1) of 0.72 W.u. Instead, in our model, the compositions of the 13/2$^+$ and 11/2$^+$ states are very different (see Fig. \ref{fragmentation}), leading to a much more quenched value of the B(M1). 

In summary, for the first time, excited states in $^{133}$Sb were observed above the 21/2$^+$, 16.6 $\mu$s isomer, up to (25/2$^+$), and lifetimes of the yrast excitations 13/2$^+$ and 15/2$^+$ were measured. To describe the structure of $^{133}$Sb, a microscopic model named ´´Hybrid Configuration Mixing Model" (HCM) was developed, which includes couplings with various types of core excitations. The model reproduces very well the energies of the observed excitations in $^{133}$Sb and provides explanation for the large differences in M1 strength for transitions connecting neighbouring medium spin yrast states. The HCM model is, therefore, a very promising tool for describing low-lying spectra of odd nuclei made of a magic core and an unpaired nucleon. The present work calls for complementary studies with direct reactions to assess spectroscopic factors of single particle states of $^{133}$Sb, in analogy with the case of the one-valence-neutron nucleus $^{133}$Sn \cite{All14}. 




The authors thank the technical services of the ILL, LPSC and GANIL for supporting the EXILL campaign. The EXOGAM collaboration and the INFN Legnaro are acknowledged for the loan of Ge detectors. This work was supported by the Italian Istituto Nazionale di Fisica Nucleare, by the Polish National Science Centre under Contract No. 2014/14/M/ST2/00738 and 2013/08/M/ST2/00257, by the UK Science and Technology Facilities Council and the UK National Measurement Office. Supports from the German BMBF (contract No. 05P12RDNUP), the Spanish MINECO FPA2013-41267-P and
NuPNET-FATIMA (PRI-PIMNUP-2011-1338) are also acknowledged.

\end{document}